\documentclass[12pt]{article}
\pdfoutput=1
\usepackage{amsmath, amsfonts, amsbsy, amsthm}
\usepackage{graphicx,psfrag,epsf}
\usepackage{enumerate}
\usepackage[sort, numbers]{natbib}
\usepackage[affil-sl]{authblk}
\usepackage{url} 
\usepackage[margin=10pt,font=small,labelfont=bf,
labelsep=endash]{caption}
\usepackage[usenames, dvipsnames]{color}

\newcommand{\bm}{\boldsymbol}

\newcommand{\bmtheta}{\boldsymbol{\theta}}
\newcommand{\bmdelta}{\boldsymbol{\delta}}

\newcommand{\bmSigma}{\boldsymbol{\Sigma}}

\newcommand{\argmin}{\argmin{\rm arg \, min}}

\newcommand{\g}{\mathbf{g}}

\newcommand{\M}{\mathbf{M}}
\newcommand{\f}{\boldsymbol{f}}
\newcommand{\1}{\mathbf{1}}
\newcommand{\E}{\mathbb{E}}
\newcommand{\tr}{\text{tr}}
\newcommand{\diag}{\text{diag}}
\newcommand{\calB}{\mathcal{B}}

\theoremstyle{definition}
\newtheorem{example}{Example}[section]

\begin{document}

\def\spacingset#1{\renewcommand{\baselinestretch}%
{#1}\small\normalsize} \spacingset{1}


\title{\bf Prioritized Data Compression using Wavelets}

\author[1,2]{Henry Scharf\thanks{Correspondence author. Email: henry.scharf@colostate.edu}}
\author[2]{Ryan Elmore}
\author[2]{Kenny Gruchalla}
\affil[1]{Department of Statistics, Colorado State University}
\affil[2]{Computational Science Center, National Renewable Energy Laboratory}
  \maketitle
\begin{abstract}
  The volume of data and the velocity with which it is being generated by
  computational experiments on high performance computing (HPC) systems is
  quickly outpacing our ability to effectively store this information in its
  full fidelity. Therefore, it is critically important to identify and study
  compression methodologies that retain as much information as possible,
  particularly in the most salient regions of the simulation space.  In this
  paper, we cast this in terms of a general decision-theoretic problem and
  discuss a wavelet-based compression strategy for its solution.  We provide a
  heuristic argument as justification and illustrate our methodology on several
  examples. Finally, we will discuss how our proposed methodology may be
  utilized in an HPC environment on large-scale computational experiments.
\end{abstract}

\noindent%

\newpage
\section{Introduction}
\label{sec:intro}

The US Department of Energy recently published a document highlighting several
issues related to data-intensive computing on future high-performance computing
(HPC) systems \citep{chen2013synergistic} with particular emphasis given to data
analysis and visualization in Chapter 4.  The authors highlight the growing
disparity between I/O and storage capabilities, and computational capabilities.
They warn that ``our ability to produce data is rapidly outstripping our ability
to use it'', particularly in a meaningful manner. This statement echoes the
sentiments expressed in previous DOE publications, {\em e.g.}
\citet{ahern2011scientific}, \citet{ashby2010opportunities}, among others. The
problem has recently manifested itself at the National Renewable Energy
Laboratory in the form of large-scale wind-turbine array simulations.  That is,
data analysis tools and the computational machinery that supports them have not
been able to scale with the HPC systems that are generating the wind-turbine
array simulations.

These considerations motivate the following research on what we term a
prioritized wavelet-based data compression methodology.  We propose storing data
in varying fidelities within the simulation space based on regions of
saliency. That is, salient regions will be stored in a high fidelity whereas the
less important regions will be more compressed.  As an example, one might
imagine the area directly in the wake of a wind turbine as being more important
in future analysis/visualization experiments and, thus, we would like to retain
the simulated data in its fullest fidelity in these regions.  On the other hand,
data on the periphery of the wind farm (less turbulent) might be less
interesting from a subsequent analytic perspective, and may be stored in a lower
fidelity in order to save space.

The wavelet representation of a signal has a history of use in both data
compression and denoising \cite{nayson_g._p._wavelets_2008}. The two
applications use the same basic algorithm of (1) performing a discrete wavelet
transform, (2) setting all coefficients in the representation whose magnitude is
below a threshold to zero, and then (3) reconstructing the signal based on the
sparse wavelet representation. Wavelet compression has most commonly been used
for images \cite{skodras_jpeg_2001} and time-series such as electrocardiogram
signals \cite{hilton_wavelet_1997}.

Optimal data compression sensitive to secondary analysis has not been generally
investigated, though there are some specific applications in image
processing. While not framed as an explicit secondary analysis, there have been
algorithms created to find optimal compression of an image sensitive to human
visual perception (see for example \cite{chandler_dynamic_2005} or the JPEG-2000
standards \cite{skodras_jpeg_2001}). We presume in our methodology that the form
of secondary analysis can be expressed explicitly as a mathematical function on
the data, however we expect the approach taken here may be extended to include
more loosely defined secondary analyses.

The use of wavelet-based compression schemes is becoming increasingly popular
in the data visualization domain, see for example \citet{gruchalla-2009},
\citet{gruchalla2009visualization}, and \citet{gruchalla2011segmentation}.  We
fully expect this trend to continue with their inclusion as the default
compression tool in the VAPOR software package \citep{clyne2010vapor}.  One of
our aims with this research is to develop a compression strategy that allows for
heterogeneous levels of compression throughout the simulation domain while
remaining consistent with VAPOR's use of the discrete wavelet transform.

While our current work does not address the motivating problem {\em per se} (
compression strategies for exascale-type problems), our intentionally narrow
focus provides the foundation upon which future research may be built.  We lay
out the mathematical background, our current problem of interest, and provide a
heuristic justification for our proposed solution in Section
\ref{sec:notation}. We illustrate the novel approach on several examples in
Section \ref{sec:examples}.  Finally, in Section \ref{sec:conc}, we summarize
our results and discuss future research directions as they relate to problems in
HPC environments.


\section{Notation/Formulation}
\label{sec:notation}
\subsection{Brief Introduction to Wavelets}
The notion of a wavelet is suggested by the name. They are `little waves', in
the sense that they possess some quality of oscillation, but have small,
localized support. A single wavelet is one member of a complete set of basis
functions with which we can represent a time series or, in general, a
function. Though there are many such sets of basis functions which qualify as
wavelets, it is useful to consider a particular set widely considered to be the
simplest. The wavelets which make up this set are called Haar wavelets, and they
are defined by translations and dilations of the so called \textit{mother} Haar
wavelet defined as
\begin{align*}
  \psi(t) &=
  \begin{cases}
    1 \; &t \in \left[0, \frac{1}{2}\right)\\
    -1 &t \in \left[\frac{1}{2}, 1\right)\\
    0 & \text{otherwise}.
  \end{cases}
\end{align*}
By choosing $j$ and $k$ appropriately, we can build a complete basis which spans
$L^2(\mathbb{R})$, and thus represent all square integrable functions as linear
combinations of these wavelets
\begin{align*}
  \psi_{j, k}(t) &= 2^{j/2}\psi(2^j t - k)
\end{align*}
with time series representation
\begin{align*}
  f(t) &= \sum_j\sum_k a_{j, k}\psi_{j, k}(t).
\end{align*}
These equations describe the continuous wavelet transform, but there are
analogous forms for discrete representations as well. For a more thorough
introduction to wavelets, see \cite{nayson_g._p._wavelets_2008}.

These basis functions are useful for many reasons, including the fact that the
wavelet representation is relatively efficient to compute, and the
representation is robust in the presence of discontinuities compared to related
methods. Additionally, the Haar wavelet is a member of a class of wavelets which
form an orthonormal basis and provides the following useful form
\begin{align}
  \sum_{t=1}^n f^2(t) &=
  \sum_k \sum_j a^2_{j, k}.
\label{eqn:parseval}
\end{align}
This identity, sometimes referred to as Parseval's relation, serves as the
foundation for a class of estimators $\hat{\f_B}$ that we propose in Section
\ref{sec:f_hat_B}.

\subsection{Decision-Theoretic Prioritization}
\label{sec:decision_theoretic}
This formulation considers the situation where the secondary analysis to be
performed upon $\f=\left(f(1), \dots, f(n)\right)^\prime$, some finite vector of
data indexed by $t=1, \dots, n$, may be explicitly expressed as a transformation
$\g\!:\!  \mathbb{R}^n \rightarrow \mathbb{R}^m$. The flexible nature of $\g$
gives this procedure a wide range of applicability. Take for example the
following hypothetical situation.

\begin{example}[Moments]
	\label{ex:moments}
Suppose we have data $\f$ for which we want to estimate the first
four moments. Our secondary function then is given by the following map from
$\mathbb{R}^n$ to $\mathbb{R}^4$:
\begin{align*}
  \g(\f) =
  \g(f(1), \dots, f(n)) &=
  \left(
    \sum_{t=1}^nf(t),\; \sum_{t=1}^nf^2(t),\;
    \sum_{t=1}^nf^3(t),\; \sum_{t=1}^nf^4(t)
  \right)^\prime.
\end{align*}
We will use the notation $g_i(\f)$ to refer to a single component of the vector
$\g(\f)$, for example
\begin{align*}
g_i(\f) = \sum_{t=1}^nf^i(t), \quad i=1, \dots, 4.
\end{align*}
We will revisit this particular example in Example \ref{ex:mom} in Section
\ref{sec:examples}.
\end{example}

We suppose now that storing the full-fidelity data $\f$ is impractical, and we
will instead be forced to make do with $\hat{\f}$, a compressed version of $\f$,
which for the moment need not be wavelet-based. In order to give some simple but
meaningful measure to the amount of error in $\hat{\f}$, and $\g(\hat{\f})$, we
propose modeling the errors $\hat{\f} - \f$ as a random vector generated from
some distribution with mean $\mathbf{0}$ and covariance matrix $\bmSigma$. In
fact, for the case of wavelet-based compression, each $\hat{f}(t) - f(t)$ will not
be random, but entirely deterministic. For even moderately large $n$ though,
these values may behave similarly to random variables. By modeling the errors
in this way, we are able to develop this problem from a decision-theoretic
perspective \cite{casella2002} and minimize the expected squared distance
between $\bmtheta = \g(\f)$ and a candidate estimator $\hat{\bmtheta} =
\g(\hat{\f})$. Specifically, we use the squared error loss function
\begin{align}
	\label{eqn:loss}
  L(\bmtheta, \bm{a}) &=
  \left(
    \bmtheta - \bm{a}\right)^\prime
    \left(\bmtheta - \bm{a}\right) =
  ||\bmtheta - \bm{a}||_2^2
\end{align}
with corresponding risk for a candidate estimator $\bmdelta$ defined by
\begin{align}
  R(\bmtheta, \bmdelta) & =
  \mathbb{E}\left[L(\bmtheta, \bmdelta)\right]. \nonumber
\end{align}
Note that we are operating under the constraint that all candidate
approximations, $\hat{\f}$, must be of the same fixed size, where
size will be a measure of the total cost of storing the approximation
$\hat{\f}$. In the case of wavelet-based compression, this is defined to be the
number of non-zero wavelet coefficients. For complicated $\g$, $\f$, and
approximations $\hat{\f}$, it may be an extremely intensive or even impossible
computation to find the optimal such $\hat{\f}$. We therefore impose two
limitations. First, we will require that we be able to linearize $\g$ and use
the first-order Taylor approximation:
\begin{align}
  \g(\hat{\f}) - \g(\f) &\approx J_{\g} \cdot (\hat{\f} - \f). \nonumber
\end{align}
where $J_{\g}$ is the $m \times n$ Jacobian matrix whose $i, t$ element is
$\partial g_i(\f)/\partial f(t)$, and $i=1, \dots, m$, $t=1, \dots,
n$. Second, we limit ourselves to a subset of the class of all fixed-size
$\hat{\f}$, defined in Section \ref{sec:f_hat_B}. This class is defined such
that approximations are relatively easy to compute, but still flexible enough in
their structure to take into account the demands of the specific secondary
analysis $\g$.

\subsection{A class of `magnifying glass' approximations $\hat{\f}_B$}
\label{sec:f_hat_B}
\begin{figure}[h!]
  \begin{center}
    \includegraphics[width=\linewidth]{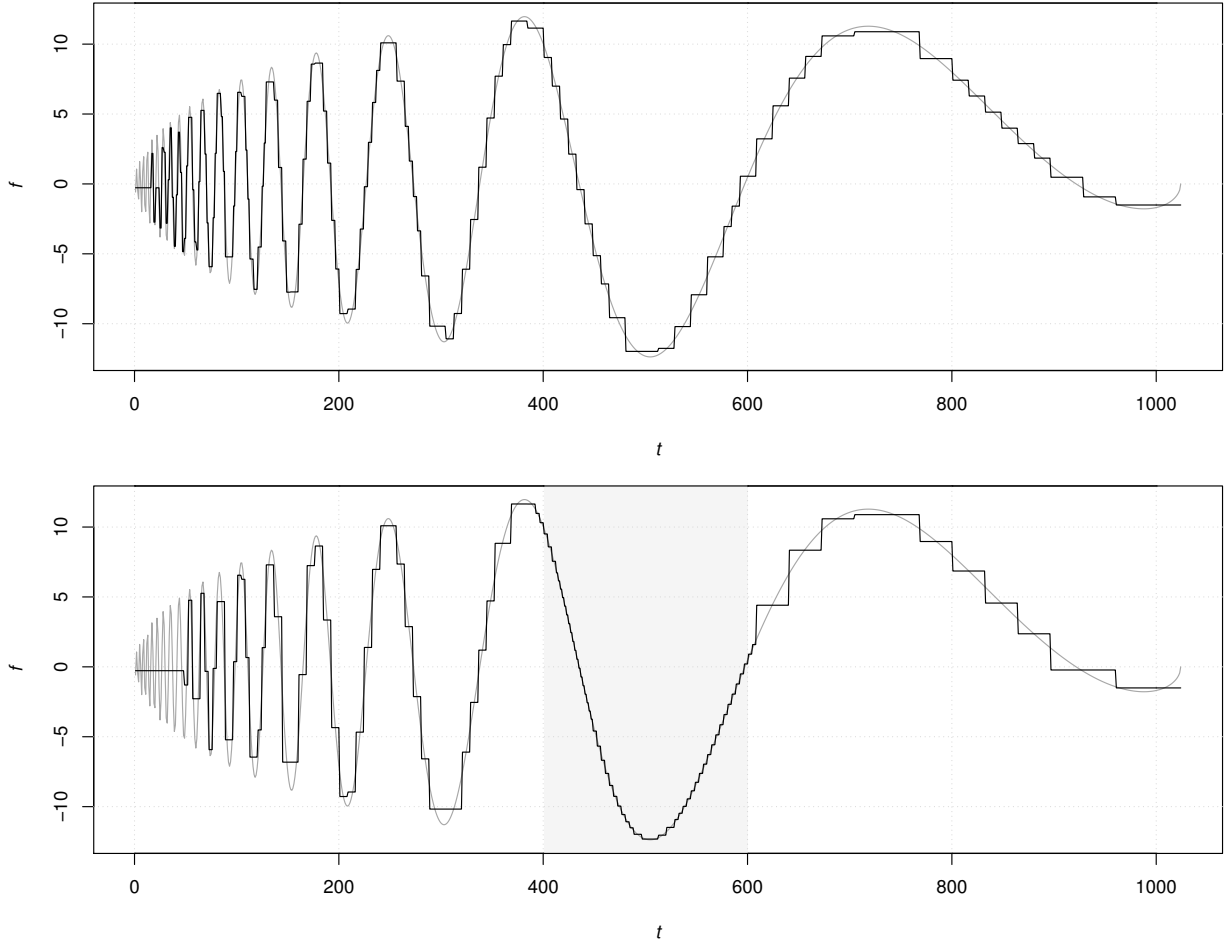}
  \end{center}
  \caption{The top plot shows the full fidelity time series in gray (`Doppler')
    with the baseline wavelet-based approximation in black. This approximation
    ignores the secondary analysis. The bottom plot shows the same time series
    in gray with an approximation $\hat{\f}_B$ in black. In this example,
    $\kappa$ was set at 0.02, with realized value $\hat{\kappa}=0.0209$, and
    $B=\{401, \dots, 600\}$ (inducated in both plots by the light shading). We
    set the size of this approximation so that only 10\% of the wavelet
    coefficients were non-zero. The `magnifying glass' approximation is
    noticeably closer to the truth in $B$ than in $B^c$.}
    \label{fig:f_hat_B_sample}
\end{figure}

From now on, $\hat{\f}$ will refer to an estimator made through (1) generating a
wavelet transform of $\f$, (2) setting a fixed number of coefficients to zero,
and then (3) reconstructing $\hat{\f}$ with this sparse wavelet
representation. The class we propose yields estimators which are practically
straightforward to produce, are mathematically tractable, and yield estimators
with significant improvements compared to a natural baseline estimator
$\tilde{\f}$. The baseline estimator is what would be produced if we completely
ignored the secondary function $\g$, and instead followed the wavelet
compression procedure which minimized the squared error norm $||\tilde{\f} -
\f||^2_2$ (where $\tilde{\f}$ is taken to be the same size as $\hat{\f}_B$). See
the top plot in Figure \ref{fig:f_hat_B_sample} for one example.

First, we write the general $\hat{\f}$ with errors partitioned into two disjoint
sets $B$ and $B^c$ whose union is the indexing set $T=\left\{1, \dots,
  n\right\}$. The $t^{th}$ element may be written as
\begin{align}
\hat{f}_{B}(t) &= f(t) + \varepsilon_B(t)\mathbf{1}_B(t) +
  \varepsilon_{B^c}(t)\mathbf{1}_{B^c}(t) \nonumber
\end{align}
where $\mathbf{1}_B(t)$ is the usual indicator function equal to one if $t \in
B$ and zero otherwise. Our class of $\hat{\f}_B$ will be those for which the
mean squared error in $B$ is proportional to the mean squared error in $B^c$.
That is, we define
\begin{align}
  \sigma^2_B &:=
  |B|^{-1} \sum_{t \in B}\varepsilon_B^2(t), \ \text{and} \nonumber \\
  \sigma^2_{B^c} &=
  \frac{1}{\kappa} \sigma^2_{B} \label{eqn:kappa}
\end{align}
where $|B| = $ the number of elements in $B$.

We can easily generate $\hat{\f}_B$ such that (\ref{eqn:kappa}) holds
(to a high level of precision) when we have orthonormal wavelets, and the set
$B$ is made up of a small number of connected intervals in $T$. The reason we
need $B$ to be this sort of set is so that we can make use of the compact
support of our wavelets. In our toy examples we only use the Haar wavelet, but
the procedure by which we generate $\hat{\f}_B$ may be extended to any
orthonormal wavelet basis with minor modifications. We use the localized nature
of the Haar wavelet basis functions along with (\ref{eqn:parseval}) in
the following way.

First, we define a set $\calB$, which is a set indexing all wavelet basis
functions whose support overlaps the region $B \in T$. We then order from
smallest to largest the coefficients of the wavelets in $\calB$ and ${\calB}^c$
by their squared value
\begin{align*}
  (a^2_{\calB (1)}, \dots, a^2_{\calB (n_J)}), \text{ and }
  (a^2_{\calB^c (1)}, \dots, a^2_{\calB^c (n_L)})
\end{align*}
and choose the unique pair of threshold values $\alpha^2_{\calB}$ and
$\alpha^2_{\calB^c}$ so that
\begin{align}
  n^{-1}_J \sum_{a^2_{\calB (j)}<\alpha^2_{\calB}}a^2_{\calB (j)} &=
  \kappa n^{-1}_L \sum_{a^2_{\calB^c (l)}<\alpha^2_{\calB^c}}a^2_{\calB^b (l)}.
  \label{eqn:energy}
\end{align}
The values $\alpha^2_{\calB}$ and $\alpha^2_{\calB^c}$ are unique because we
have fixed the number of non-zero coefficients.

Because of the localized support of the wavelets, when $B$ is made up of a small
number of connected subsets of $T$, the number of wavelets with support in both
$B$ and $B^c$ will be small compared to the number of wavelets with support
entirely in $B$ or $B^c$. Moreover, the coefficients which typically have small
magnitudes are those which correspond to basis functions at finer scales, which
means the wavelet coefficients which are most likely to be effected by
thresholding are the ones who tend to have support entirely in $B$ or
$B^c$. Therefore we will have $|B| \approx n_J$ and also
\begin{align*}
  \frac1{|B|}\sum_{t \in B}(\hat{f}_{B}(t) - f(t))^2 &\approx
  \frac1{n_J}\sum_{a^2_{\calB(j)}<\alpha^2_{\calB}}a^2_{\calB (j)}
\end{align*}
as well as the analogous result for $B^c$ and $\calB^c$. Therefore, ensuring
(\ref{eqn:energy}) in turn ensures that (\ref{eqn:kappa})
approximately holds. In practice, when we implemented this procedure, we were
able to generate $\hat{\f}_B$ for which $\hat{\kappa}$, the realized value of
the ratio of mean squared errors, came very close to $\kappa$. When we specified
$\kappa = 0.1$, for example, our realized ratio was generally in $(0.09, 0.11)$.

These $\hat{\f}_B$ are in some sense the simplest possible way to take into
account the demands of the secondary analysis. We are partitioning the indexing
set $T$ into two subsets, where values $f(t)$, $t \in B$ are \textit{more
  important} than values $f(t^*)$, $t^* \notin B$ in accurately estimating
$\g(\f)$ by a factor of $\kappa$. Therefore, a datum $f(t)$ is either important
or \textit{not} important, there is no spectrum of importance. Familiarly, we
call these approximations $\hat{\f}_B$ `magnifying glass' estimators with
magnification factor $\kappa^{-1}$ since they effectively give us a closer look
at region $B$ compared to $B^c$ (see bottom plot of Figure
\ref{fig:f_hat_B_sample}).  More subtle schemes are certainly worth
investigating, but this first-order approach already yields promising results.
Figure \ref{fig:f_hat_B_sample} shows an implementation of this method. In the
top plot, the gray curve shows the full-fidelity time series (`Doppler'
\cite{nayson_g._p._wavelets_2008}) and the black step function is the baseline
approximation $\tilde{\f}$. In the bottom plot, the gray curve is the same, but
the approximation shown is a magnifying glass estimator with $B=\{401, \dots,
600\}$ and $\kappa=0.02$ ($\hat{\kappa}=0.0209$). The size of the
approximation is such that 10\% of the wavelet coefficients were allowed to be
non-zero. The higher fidelity region appears to extend slightly beyond $B$ on
both ends because of the presence of wavelet basis functions with overlapping
support.

The values $\sigma^2_B$ and $\sigma^2_{B^c}$ will depend both on $\kappa$ and
the size of the approximation. For instance, the smallest possible $\sigma^2_B$
and $\sigma^2_{B^c}$ for a given approximation size will occur when $\kappa=1$.
As $\kappa$ decreases toward 0, we are sacrificing some overall increase in the
error in $\hat{\f}_B$ in exchange for better estimating the parts of $\f$ that
are most important for the secondary analysis $\g$, resulting in an overall
improvement in the precision of $\g(\hat{\f}_B)$.

\subsection{Importance Function}
\label{sec:importance_function}
In order to add some degree of saliency to regions within the simulation
space, we introduce the `importance function' concept.  Recall the loss function
defined in \ref{eqn:loss} and note that it can be re-written
(approximately) as
\begin{align*}
  L(\g, \hat{\f}) &\approx
  \left[J_{\g} \cdot (\hat{\f} - \f)\right]^\prime
  \left[J_{\g} \cdot (\hat{\f} - \f)\right]
  \\ \nonumber
  &= (\hat{\f} - \f)^\prime J_{\g}^\prime J_{\g} (\hat{\f} - \f)\\ \nonumber
  &= (\hat{\f} - \f)^\prime \M (\hat{\f} - \f).
\end{align*}
We will refer to $\M:=J_{\g}^\prime J_{\g}$ as the `importance matrix' and its
diagonal elements as the `importance function'. This name will become clear
later when we show that the diagonal of this matrix largely determines the form
of the optimal approximation. These $n$ diagonal elements can be thought of as
defining a level of importance or weight for each datum $f(t)$.

We will now model the deterministic errors $\hat{\f} - \f$ as a random vector
with mean vector $\mathbf{0}$ and covariance matrix $\bmSigma$. For
approximations of type $\hat{\f}_B$, a natural covariance matrix to ascribe to
this random variable is a diagonal matrix with elements equal to either
$\kappa\sigma^2_{B^c}$ or $\sigma^2_{B^c}$. Using this model, our loss function
takes a quadratic form, and the risk is approximated by
\begin{align}
  R(\g, \hat{\f}_B) &\approx
  \E\left[(\hat{\f} - \f)^T\M(\hat{\f} - \f)\right] \nonumber\\
  &=\tr(\M\Sigma) \nonumber\\
  &=\sigma^2_{B^c}\left[\kappa\sum_{t \in
      B}\sum_{i=1}^m \left(\frac{\partial g_i}{\partial f(t)}\right)^2
    + \sum_{t \notin B}\sum_{i=1}^m \left(\frac{\partial g_i}{\partial
        f(t)}\right)^2\right]. \label{eqn:risk}
\end{align}
In order to minimize $R(\g, \hat{\f}_B)$, we will need to take into account two
effects determined by $B$. First, as the size of $B$ increases, $\sigma^2_{B^c}$
will increase, since more and more error in $B^c$ will be sacrificed to maintain
the higher fidelity in $B$. Second, the choice of $B$ will have an impact on the
two sums. If we choose $B$ such that the largest values of $\sum_{i=1}^m
\left(\partial g_i/\partial f(t)\right)^2$ are in the first term of Equation
(\ref{eqn:risk}), then we will reduce our risk function $R(\g, \hat{\f}_B)$
because $\kappa<1$. Taking these together, we can intuitively expect that the
optimal set $B$ will be the smallest possible set such that the large elements
on the diagonal of $\M$ have index inside $B$.

\subsection{Idealized Secondary Analysis} 
\label{sec:contrived}
We now turn our attention to one idealized setting where the secondary analysis
takes a contrived form as a basic check on casting the problem in this
decision-theoretic manner. Consider the following secondary calculation $\g$
defined by
\begin{align*}
  \g(\f)_i & := f(i)(\lambda\1_A(i) + \1_{A^c}(i)), \quad i = 1, \dots, n
\end{align*}
or written more explicitly
\begin{align*}
  g_i(\f) &=
  \begin{cases}
    \lambda f(i), \quad & i\in A\\
    f(i), & i \notin A
  \end{cases} \\
  i &\in \{1, \dots, 1024\}.
\end{align*}
This represents the case where our secondary analysis consists merely of
multiplying some elements of $\f$ by a scalar $\lambda$, and retaining the rest
unaltered. This is akin to placing some region of the data under a magnifying
glass with a magnification factor of $\lambda$, and we therefore expect an
estimator of type $\hat{\f}_B$ to be a good choice. For $\lambda$ greater than
one, we expect that the optimal approximation $\hat{\f}_B$ ought to be one where
$B \approx A$, since this is clearly the `important' part of $\f$. The
importance matrix is straightforward to calculate, since it is the product of
two diagonal matrices with elements equal to either $\lambda$ or 1 depending on
the index's membership in $A$.

In the argument that follows, we let $\lambda$ go to infinity, and $\kappa$ go
to zero, which corresponds to the case where $\g$ and $\f$ show infinite
preference for region $A$ and $B$ respectively. In this extreme we will see that
it is possible to verify $B=A$.

\begin{figure}[h!]
  \begin{center}
    \includegraphics[width=\linewidth]{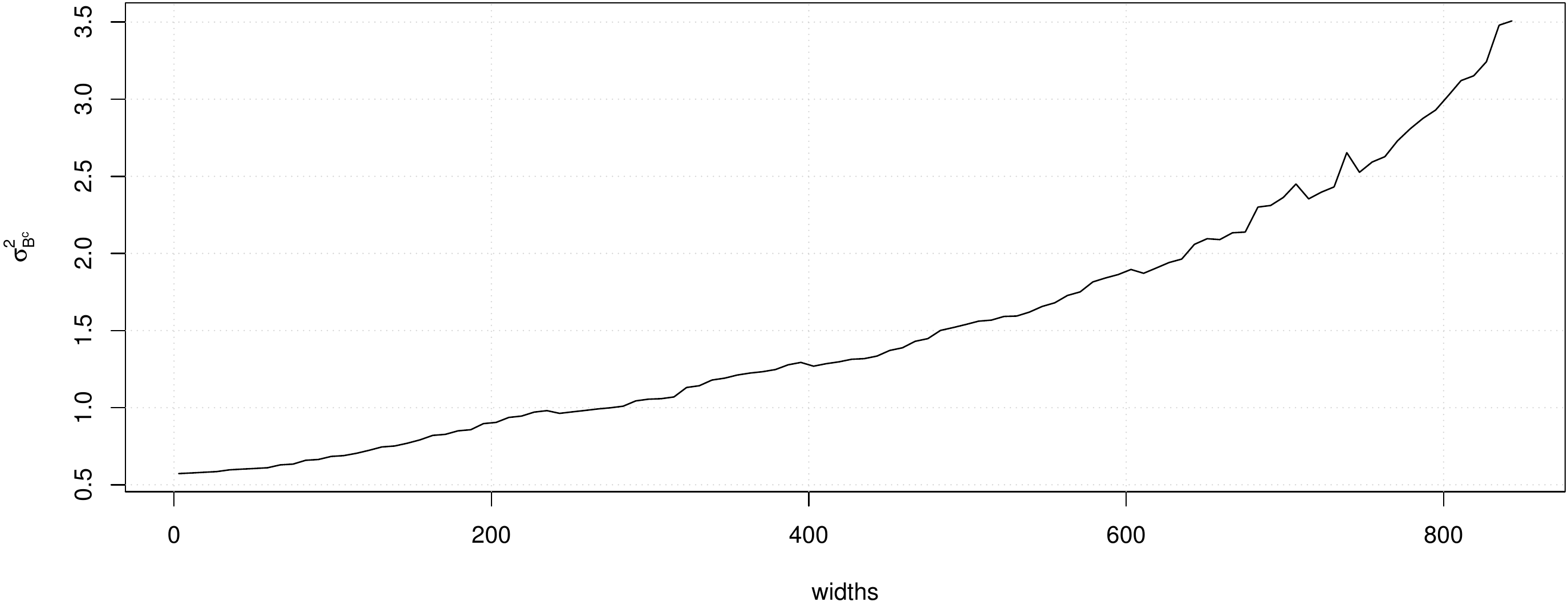}
  \end{center}
  \caption{This plot shows the roughly monotone behavior of $\sigma^2_{B^c}$ as
    we increase the size of $B$ for fixed $\kappa=0.1$. In this case, we used
    the `Doppler' time series and let $B$ increase from $\{511, 512, 513\}$ to
    $\{91, \dots, 931\}$, increasing the width by 4 in each direction at each
    step. The units on the $y$-axis are specific to this example, but the monotone
    behavior is generally applicable.}
    \label{fig:sigmaB_vs_sizeB}
\end{figure}

Starting from (\ref{eqn:risk}) we have
\begin{align*}
  R(\g, \hat{\f}_B) &=
  \sigma^2_{B^c}\left[
    \kappa\left(
      \sum_{t \in A \cap B}\lambda^2 +
      \sum_{t \in A^c \cap B}1\right) +
    \left(
      \sum_{t \in A \cap B^c}\lambda^2 +
      \sum_{t \in A^c \cap B^c}1\right)
  \right]\\
  &= \sigma^2_{B^c}\left[
    \kappa\left(
      \lambda^2|A \cap B| + |A^c \cap B|\right) +
    \lambda^2|A \cap B^c| + |A^c \cap B^C|
  \right].
\end{align*}
As $\lambda$ approaches infinity, the risk will also become infinite, so we
next normalize by the constant $\lambda^2$ to get
\begin{align}
  \frac{R(\g, \hat{\f}_B)}{\lambda^2} &=
  \sigma^2_{B^c}\left[
    \kappa\left(
      |A \cap B| + \frac{|A^c \cap B|}{\lambda^2}\right) +
    |A \cap B^c| + \frac{|A^c \cap B^c|}{\lambda^2}
  \right]. \label{eqn:normrisk}
\end{align}
Minimizing (\ref{eqn:normrisk}) is equivalent to minimizing the risk, but now as
we let $\lambda^2$ grow large, the quantity of interest simplifies to
\begin{align}
  \lim_{\lambda^2 \rightarrow \infty}
  \left(
    \frac{R(\g, \hat{\f}_B)}{\lambda^2}
  \right) &=
  \sigma^2_{B^c}
  \left(
    \kappa |A \cap B| + |A \cap B^c|
  \right). \label{eqn:limnormrisk}
\end{align}
To get any further, we need to know more about how $\sigma^2_{B^c}$ depends on
our choice of $B$. In practice, this relationship will be complicated, but
roughly speaking we expect $\sigma^2_{B^c}$ to be monotone increasing as $|B|$
increases as discussed in Section \ref{sec:importance_function} (see Figure
\ref{fig:sigmaB_vs_sizeB}). From (\ref{eqn:limnormrisk}), we can see that the
second factor in the limiting risk is independent of the size of $A^c \cap B$,
and so we might as well choose $B^\prime = A \cap B$ instead of $B$. That is,
\begin{align*}
  \lim_{\lambda^2 \rightarrow \infty}
  \left(
    \frac{R(\g, \hat{\f}_B)}{\lambda^2}
  \right) &\geq
  \lim_{\lambda^2 \rightarrow \infty}
  \left(
    \frac{R(\g, \hat{\f}_{A \cap B})}{\lambda^2}
  \right),
\end{align*}
since $\sigma^2_{B^c}$ is monotone increasing in $|B|$. It is clear the optimal
$B$ will be contained within $A$.

Letting $\kappa$ go to zero now yields
\begin{align*}
  \lim_{\lambda^2 \rightarrow \infty, \; \kappa \rightarrow 0}
  \left(
    \frac{R(\g, \hat{\f}_B)}{\lambda^2}
  \right) &=
  \sigma^2_{B^c}|A \cap B^c|.
\end{align*}
Now as $B$ increases in size to approach $B=A$, $\sigma^2_{B^c}$ will increase,
but when $B=A$, the second term is zero and the normalized risk vanishes. We
will see in Examples \ref{ex:ind_dop} and \ref{ex:ind_bumps} that even when
$\lambda$ and $\kappa$ are far from infinity and zero, this result still holds
to large degree.

\section{Examples}
\label{sec:examples}
In considering possible sets $B \subset T$, it is computationally intractable to
consider all $2^{|T|}$ possible subsets. Often though, we may have reason to
believe \textit{a priori} that the most important parts of $\f$ lie
predominantly in a few connected subsets of the space $T$. For instance, if the
data are spatially or temporally arranged, we may expect strong positive
correlation in importance between neighboring data. In our motivating example
related to wind turbine arrays, we expect that the most important regions are in
and around the wakes. In the following examples we restrict the space of
possible $\hat{\f}_B$ to include only those for which $B$ is a single connected
interval in $T$.

We performed a nearly-exhaustive (see Appendix \ref{app:searches}) search for
the optimal $B$ for several toy examples. In each example our data $\f$ is one
of two different one-dimensional time series, both widely used in simulation
studies involving denoising and density estimation (\cite{donoho1994b} and
\cite{donoho1995adapting}), and also used for illustration in
\cite{nayson_g._p._wavelets_2008}. In each case we also specify a secondary
analysis $\g$, and a value for $\kappa$. In all examples, the size of the
approximation is such that 10\% of the wavelet coefficients were allowed to be
non-zero. The nearly-exhaustive search considers almost all possible
uninterrupted intervals within $T=\{1, \dots, 1024\}$ for a fixed value
$\kappa$. For each interval, the relative squared error is defined as
\begin{align*}
  SE_B&=L(\g, \hat{\f}_B), \\
  SE_0&=L(\g, \tilde{\f}), \& \\
  \mbox{relSE}&:=SE_B/SE_0.
\end{align*}
Therefore, values of $\mbox{relSE}$ less than 1 represent improvements over the
baseline wavelet compression, $\tilde{\f}$, which ignores the secondary
function. The interval corresponding to the smallest $\mbox{relSE}$ is (nearly) optimal
in the space of single uninterrupted intervals.

\begin{example}[Idealized Secondary Analysis for `Doppler' Time Series]
  \label{ex:ind_dop}

  We offer here the results of implementation on a particular example of the
  idealized case mentioned in Section \ref{sec:contrived}. For demonstration,
  the secondary function is
  \begin{align*}
    \g_1(\f)_i &= f(i)\left[
      5\times \mathbf{1}_{\{401, \dots, 600\}}(i) +
      \mathbf{1}_{\{401, \dots, 600\}^c}(i)
    \right], \quad i=1, \dots, n
  \end{align*}
  so that $\lambda = 5$ and $A = \{401, \dots, 600\} \subset T=\{1, \dots,
  1024\}$. We therefore expect that the optimal $\hat{\f}_B$ will occur when $B
  \approx \{401, \dots, 600\}$. Figure \ref{fig:ind_dop} shows the results of a
  nearly-exhaustive search over all single intervals $B$ in $T$ with fixed
  $\kappa=1/10$ for our first time series (`Doppler'). The top plot is the full
  fidelity time series $\f$, with results overlaid. The bottom plot shows the
  importance function ($\diag(\M)$) with results overlaid. Each overlaid segment
  represents a proposed interval $B$, plotted at a height proportional to
  $\mbox{relSE}$, the ratio of squared error in $\hat{\f}_B$ to $\tilde{\f}$,
  the approximation which ignores $\g$. The axis to the right shows these
  $\mbox{relSE}$. Since this search procedure is naive, many of the proposed
  $\hat{\f}_B$ are in fact much worse than $\tilde{\f}$. We have included only
  the top performing $B$, in this case the top 2\%.

  The interval with the smallest $\mbox{relSE}$ is $\{413, \dots, 597\}$, which
  is almost exactly equal to the interval where the diagonal of $\M$ is
  large. As we saw in Figure \ref{fig:f_hat_B_sample}, our method for generating
  $\hat{\f}_B$ tends to produce an approximation to $\f$ in which the high
  fidelity region bleeds slightly beyond the specified $B$, which may explain
  why the optimal interval is slightly narrower than we expected. Additionally,
  we can see in the bottom plot that the top 2\% of intervals are all stably
  located near $\{401, \dots, 600\}$, and this can be verified for a range of
  $\kappa$ values (we verified $\kappa=(1/5, 1/10, 1/20)$. In fact, this
  stability is visible well beyond the top 2\% of $\mbox{relSE}$, but plots
  including a larger proportion of proposed segments are cluttered and difficult
  to interpret.

  \begin{figure}[h!]
    \begin{center}
      \includegraphics[width=\linewidth]{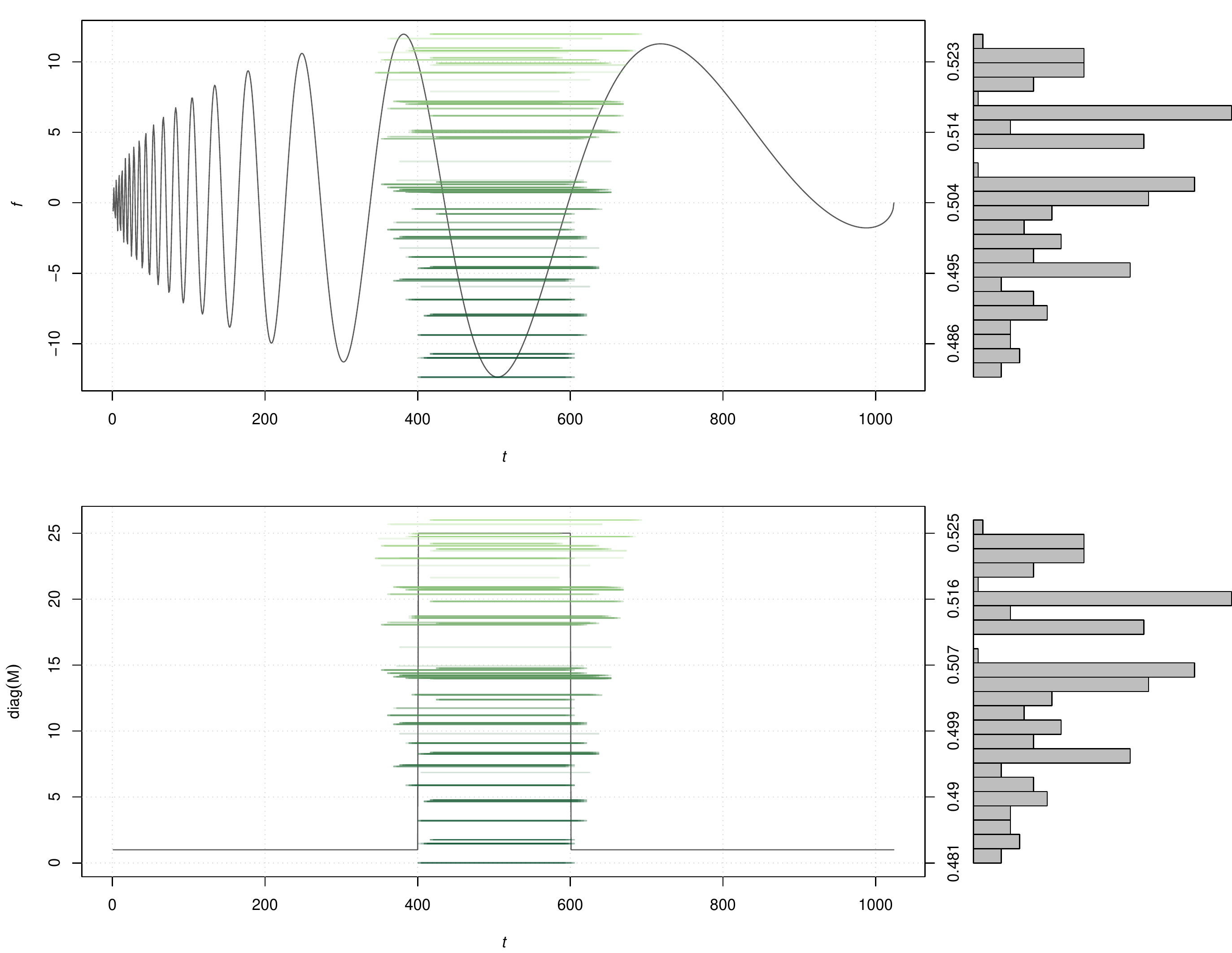}
    \end{center}
    \caption{Top 2\% of $B$ for the `Doppler' time series and secondary function
      $\g_1$ in Example \ref{ex:ind_dop}. Each horizontal line segment
      represents an interval $B$, and the height of each segment is proportional
      to the improvement over the approximation which ignores secondary
      analysis, $\tilde{\f}$. The ratio of squared error loss for $\hat{\f}_B$
      to that of $\tilde{\f}$ is shown in the right axis. The histogram to the
      right of each plot shows the distribution of the top 2\% of proposed
      intervals. Some segments $B$ have the same $\mbox{relSE}$, so the
      histogram helps to illuminate where segments have overlapped.}
    \label{fig:ind_dop}
  \end{figure}
\end{example}

\begin{example}[Idealized Secondary Analysis for `Doppler' Time Series]
  \label{ex:ind_bumps}

  In this example, we keep the same secondary analysis $\g_1$, but we examine a
  different time series called `Bumps'. Figure \ref{fig:ind_bumps} shows the
  same pair of plots as in Figure \ref{fig:ind_dop} for this new time series. It
  is interesting to note the influence here of not just the importance function,
  but also the nature of $\f$, in particular the places where the time series is
  equal to zero. In these regions, the approximation that the errors $\hat{\f} -
  \f$ are randomly distributed with variance $\sigma^2_B$ or $\sigma^2_{B^c}$ is
  unreasonable. Since it is trivial to represent data that are identically zero,
  the errors here will be zero for any reasonably sized approximation, {\em
    i.e.}, for any approximation with more than a very small number of non-zero
  wavelet coefficients), regardless of our specification of $\kappa$. The
  optimal interval is $\{405, \dots, 461\}$. As is visible in the top plot, this
  region corresponds roughly to the portion of $A$ where $\f$ is non-zero.
  \begin{figure}[h!]
    \begin{center}
      \includegraphics[width=\linewidth]{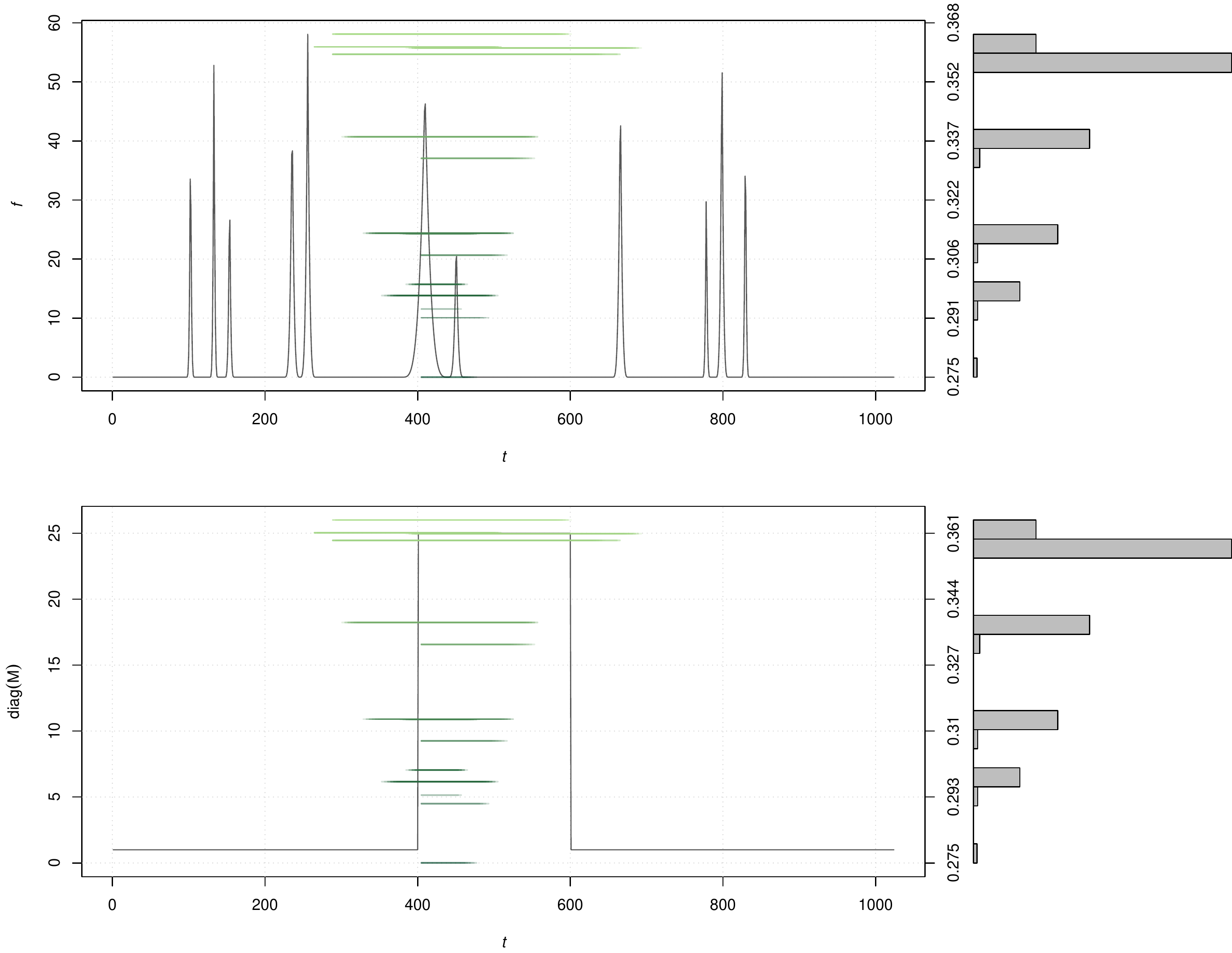}
    \end{center}
    \caption{Top 5\% of $B$ for the `Bumps' time series and secondary function
      $\g_1$ in Example \ref{ex:ind_bumps}. Each horizontal line segment
      represents an interval $B$, and the height of each segment is proportional
      to the improvement over the approximation which ignores secondary
      analysis, $\tilde{\f}$. The ratio of squared error loss for $\hat{\f}_B$
      to that of $\tilde{\f}$ is shown in the right axis. The histogram to the
      right of each plot shows the distribution of the top 2\% of proposed
      intervals. Some segments $B$ have the same $\mbox{relSE}$, so the
      histogram helps to illuminate where segments have overlapped. In this
      example, it is interesting to note the influence of not just the
      importance function $\diag(\M)$, but also the nature of the data $\f$
      themselves. In particular, there are many indices for which the data are
      close to zero. The optimal interval (most easily seen on the middle
      figure) is $\{405, \dots, 461\}$, which corresponds roughly to the portion
      of $A$ where $\f$ is non-zero.}
    \label{fig:ind_bumps}
  \end{figure}
\end{example}

\begin{example}[Exponential Sine]
  \label{ex:expsin}

  We next consider a secondary analysis for which we will have little
  intuition. The function $\g$ is still chosen to be a map to $\mathbb{R}^n$,
  but the function is more complicated than the magnifying glass situation. We
  define the secondary analysis to be
  \begin{align*}
    \g_3(\f)_i &= e^{f(i)/6}\sin\left(f(i)\right), \quad i = 1, \dots, n.
  \end{align*}
  The shape of the importance function is now something more complicated than
  the function in the previous two examples. Without looking at the importance
  function, it is difficult to guess \textit{a priori} where the most
  `important' data will be, and where the lens of our magnifying glass
  approximation should lie. In this example we reuse the `Doppler' data.

  Figure \ref{fig:expsin} shows results overlaying the data (top) and the
  importance function (bottom). The best proposed intervals $B$ cover the region
  of the importance function where there are the most large values. Though the
  importance function attains large values in the interval $\{600, \dots, 900\}$
  as well as in the interval $\{350, \dots, 425\}$, the former has more such
  values, and since we are only considering $B$ that are single intervals, it
  makes sense that the best $B$ are clustered more or less in this range.
  \begin{figure}[h!]
    \begin{center}
      \includegraphics[width=\linewidth]{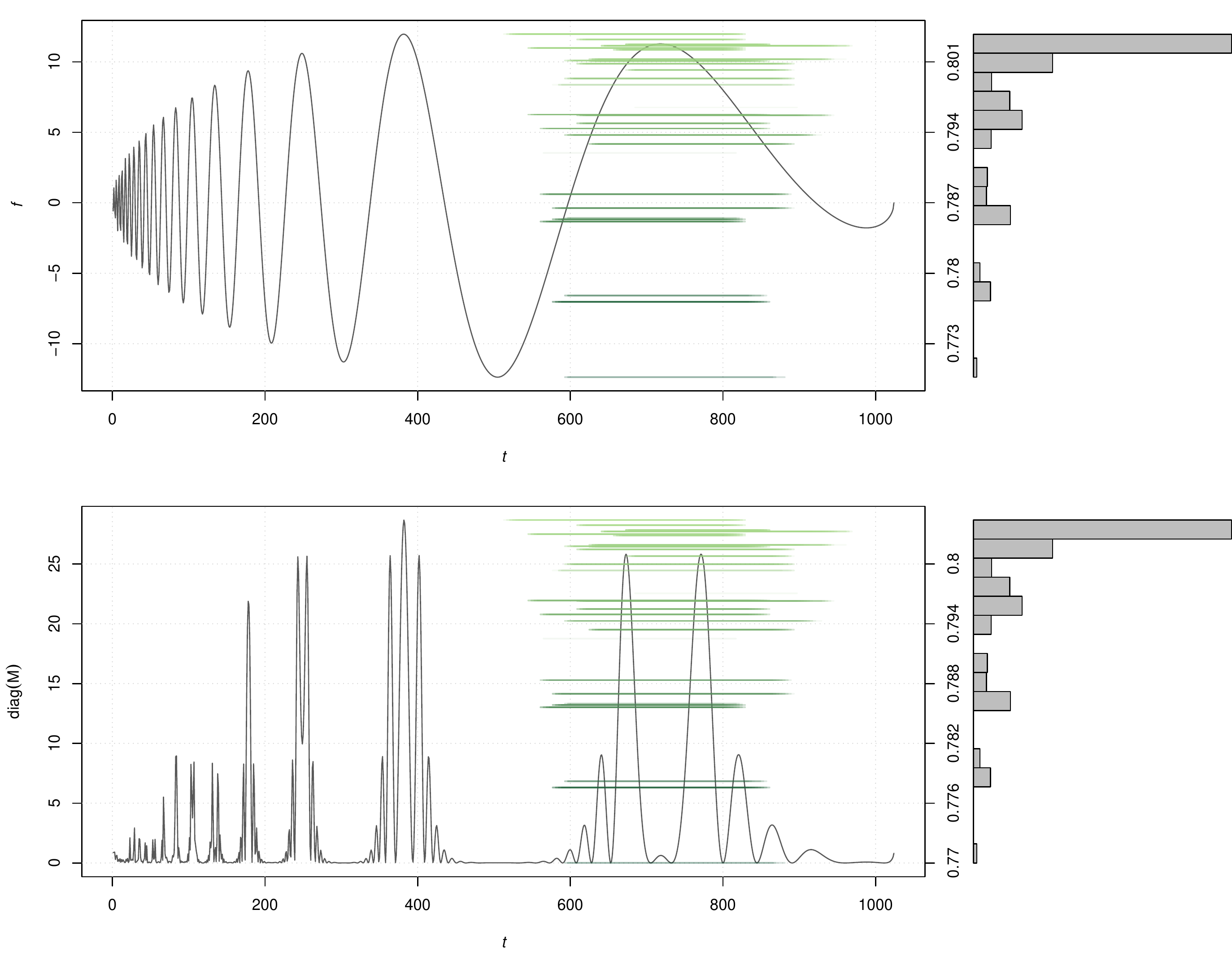}
    \end{center}
    \caption{Top 5\% of $B$ for the `Doppler' time series and secondary function
      $\g_3$ in Example \ref{ex:expsin}. Each horizontal line segment represents
      an interval $B$, and the height of each segment is proportional to the
      improvement over the approximation which ignores secondary analysis,
      $\tilde{\f}$. The ratio of squared error loss for $\hat{\f}_B$ to that of
      $\tilde{\f}$ is shown in the right axis. The histogram to the right of
      each plot shows the distribution of the top 2\% of proposed
      intervals. Some segments $B$ have the same $\mbox{relSE}$, so the
      histogram helps to illuminate where segments have overlapped.}
    \label{fig:expsin}
  \end{figure}
\end{example}

\begin{example}[Moment estimators]
  \label{ex:mom}
  We consider one final secondary analysis on the `Doppler' data, first
  mentioned in Example \ref{ex:moments}. In this example, we consider a
  secondary analysis that summarizes the raw data using the first four
  moments
  \begin{align*}
    \g_4(\f) &=
    \left(
      \sum_{t}f(t), \sum_{t}f^2(t),
      \sum_{t}f^3(t), \sum_{t}f^4(t)
    \right)^\prime.
  \end{align*}
  The corresponding Jacobian and importance functions, respectively, are
  \begin{align*}
    J_{\g_4} &=
    \begin{bmatrix}
      1 & 1 & \dots & 1\\
      2f(1) & 2f(2) & \dots & 2f(n)\\
      3f^2(1) & 3f^2(2) & \dots & 3f^2(n)\\
      4f^3(1) & 4f^3(2) & \dots & 4f^3(n)
    \end{bmatrix}
  \end{align*}
  and
  \begin{align*}
    \diag(\M)(t) &= 1 + (2f(t))^2 + (3f^2(t))^2 + (4f^3(t))^2, \quad t=1, \dots, n.
  \end{align*}
  This is our first example for which $m \neq n$, and is intended to
  be the most realistic. However, these results may be interpreted in much the
  same manner as the other examples.

  Figure \ref{fig:mom_dop} shows the results overlaying the `Doppler' time
  series (top) and the importance function (bottom). As in the previous
  examples, we can see that the intervals $B$ corresponding to the best
  magnifying glass type approximations include the portions of the importance
  function with the largest values.
  \begin{figure}[h!]
    \begin{center}
      \includegraphics[width=\linewidth]{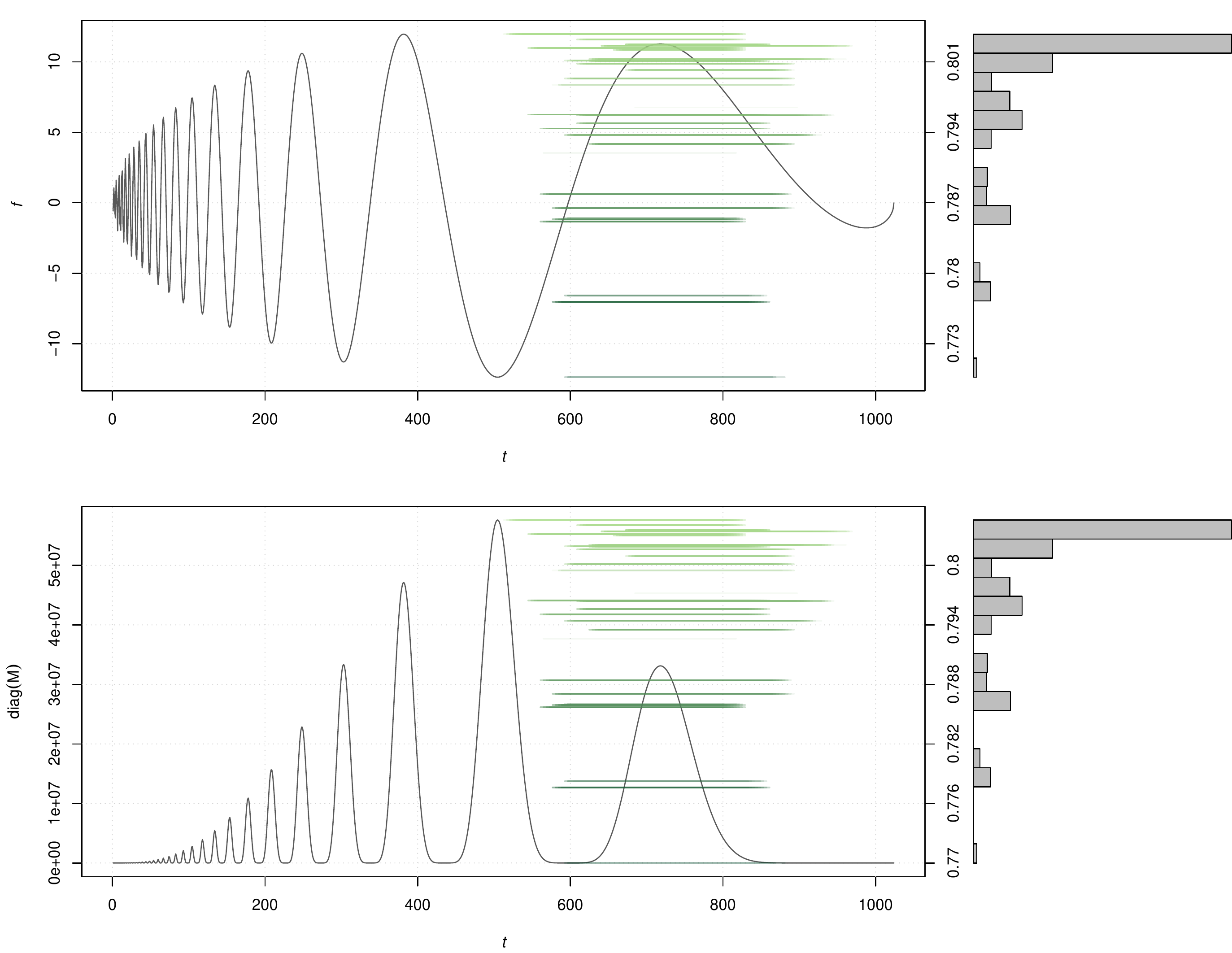}
    \end{center}
    \caption{Top 2\% of $B$ for the `Doppler' time series and secondary function
      $\g_4$ in Example \ref{ex:mom}. Each horizontal line segment represents an
      interval $B$, and the height of each segment is proportional to the
      improvement over the approximation which ignores secondary analysis,
      $\tilde{\f}$. The ratio of squared error loss for $\hat{\f}_B$ to that of
      $\tilde{\f}$ is shown in the right axis. The histogram to the right of
      each plot shows the distribution of the top 2\% of proposed
      intervals. Some segments $B$ have the same $\mbox{relSE}$, so the
      histogram helps to illuminate where segments have overlapped.}
    \label{fig:mom_dop}
  \end{figure}
\end{example}

\section{Discussion and Future Work}
\label{sec:conc}
There are many hurdles to clear in DOE's race towards exascale high performance
computing environments including future system architectures, power/energy
compliance, and many others. In this paper, we discuss one such problem related
to the overwhelming demands being placed on I/O and storage systems due to the
increased volume of data being generated by computational experiments.  We
remind the reader that our treatment of this larger-scale problem is in no way
complete. However, we do believe that our codification of a simpler class of
heterogeneous wavelet-based compression strategies will provide practitioners
and theoreticians a systematic framework for future work.  Our initial
experiments given in Section \ref{sec:examples} certainly supports this
assertion.

In closing, we will briefly discuss the obvious extension for this current
work. First, it may be unrealistic to assume that we know an importance function
{\em a priori}.  Consider our motivating example involving wind turbine arrays.
We might know the approximate locations within the simulation domain that we
would like to store in a high fidelity, however, the precise locations will be
unknown at run-time.  In other words, we will need to estimate the optimal
prioritized region without using $\g(\f)$ and an exhaustive (or near-exhaustive)
search.  A smart {\em in situ} sampling strategy of the data should enable the
development of a reasonably good estimator, $\hat{\M}$.


\bigskip
\begin{center}
{\large\bf Acknowledgements}
\end{center}
The authors would like to thank Professor Jay Breidt and David Biagioni for
their suggestions on improving early drafts of this manuscript.

This work was supported by the Laboratory Directed Research and Development
(LDRD) Program at the National Renewable Energy Laboratory.  NREL is a national
laboratory of the U.S. Department of Energy Office of Energy Efficiency and
Renewable Energy operated by the Alliance for Sustainable Energy, LLC.

\appendix
\section{Nearly Exhaustive Searches}
\label{app:searches}
All data used in the examples in this paper were of size 1024. When searching
for the optimal $B$, we were unable to consider every interval within $T$,
however we we able to consider nearly every interval, and we do not expect
substantive changes would occur in our results if we did check all
possibilities.

By nearly exhaustive, we mean the following. Instead of all intervals
\begin{align*}
  \{a, \dots, b\}, \quad 1 \leq a < b \leq n,
\end{align*}
we considered only intervals of the form where
\begin{align*}
  a &= 4x + 1, \quad &x=1, \dots, 255\\
  b &= 4y + a, &\text{for all } y = 1, \dots, 81 \text{ such that } b<1024.
\end{align*}
We therefore considered all intervals of
length equal to a multiple of 4 up to 324, and left boundary equal to all
integers with a value of 1 modulo 4 up to 1021, {\em e.g.}, we checked $513
\dots 513 + 32$ but {\em not} $514 \dots 514 + 32$. For the cases where this
system led to ranges that would extend beyond $\{1, \dots, 1024\}$, we simply
took the intersection of the proposed $B$ with $T$. The
convenience of this systematic approach outweighed the slight preference shown
for boundaries that included points near the right edge of $\{1, \dots, 1024\}$

This nearly-exhaustive method was used for convenience. The
computational demands for implementation were reasonable for a
personal laptop computer (on the order of 10 minutes per search).
%
%
%
%

\bibliographystyle{apalike}
\bibliography{PWC.bib}

\end{document}